\documentclass[twoside,10pt,a4paper]{newFNLstyle}
\input{epsf}
\usepackage{graphics,graphicx}
\usepackage{cite}

\begin{document}

\volnumpagesyear{0}{0}{000--000}{2001}
\dates{received date}{revised date}{accepted date}

\title{INVERTED REPEATS IN VIRAL GENOMES}

\authorsone{M. Span\`o, F. Lillo,  S. Miccich\`e and R. N. Mantegna}
\affiliationone{INFM Unit\`a di Palermo and Dipartimento di Fisica e Tecnologie Relative}
\mailingone{viale delle Scienze, Edificio 18, I-90128, Palermo, Italy}


\maketitle

\markboth{INVERTED REPEATS IN VIRAL GENOMES}{M. Span\`o, F. Lillo,  S. Miccich\`e and R. N. Mantegna}

\pagestyle{myheadings}


\begin{abstract}
We investigate 738 complete genomes of viruses 
to detect the presence of short inverted repeats.
The number of inverted repeats found is compared with the prediction obtained for a Bernoullian and
for a Markovian control model.
We find as a statistical regularity that the number of observed 
inverted repeats is often greater than the one expected in terms of a
Bernoullian or Markovian model in several of the viruses
and in almost all those with a genome longer than 30,000 bp.  
\end{abstract}

\section{Introduction}

In the last few years there has been a progressively growing interest
about the role of noncoding RNA (ncRNA) sequences producing functional 
RNA molecules having regulatory roles \cite{Eddy}. Prominent 
examples of these
new regulatory RNA families are microRNA (miRNA) \cite{Lagos,Lau,Lee} and
small interference RNA (siRNA) \cite{Hamilton,Hutvagner}. Most of these 
structures shares the property of being characterized by a hairpin 
secondary structure. DNA or RNA short sequences that may be associated
to RNA secondary structures are present in genomes of different
species of phages, viruses, bacteria and eukaryotes. Indication
about the potential existence of RNA secondary structures can be inferred
throughout the detection of short pair sequences having the characteristic 
of inverted repeats (IRs) in the investigated genomes \cite{Schroth}. 

In the present study we systematically investigate
all the complete genomes of viruses publicly available at 
{\tt{http://www.ncbi.nlm.nih.gov/}} on April 2003 to detect the presence of 
short IRs. The complete list containing the accession numbers of the investigated 
genome sequences is accessible at the web-page: {\tt{http://lagash.dft.}}
{\tt{unipa.it/viruses/}}{\tt{List.txt}}. 
The number of IRs found for different classes of structures
and for each set of control parameters is compared with the prediction 
obtained for a Bernoullian (i.e. indipendent and identically distributed 
nucleotide occurence) and for a Markovian control models.

With this technique we are able to evaluate the presence
of a large number of IRs that cannot be explained in terms of simple
control models therefore indicating their potential biological role.
For each virus, the study is performed -- (i) over the entire genome 
and (ii) in its coding and noncoding regions.

\section{Viral Genomes}
During the past years about heigth hundred viral genomes have been 
completely sequenced. The complete sequence of their genomes is publicly
accessible at specialized web pages. The database comprises different classes 
of viruses characterized by single stranded
or double stranded nucleic acids, different infected organisms etc.
In the present study we search the genomes for the presence of short subsequences, 
which might be associated with the existence of a secondary 
structure in regions of RNA originating from that subsequences.

Hairpin structures can occur when an IR 
is present in the nucleic acid sequence. For example, the DNA sequence 
5'aGGAATCGATCTTaacgAAGATCGATTCCa3' is a sequence having a sub-sequence
GGAATCGATCTT which is the IR of AAGATCGATTCC.
IRs of this type can form a hairpin having a stem of length 
12 nucleotides and a loop (aacg) of length 4  nucleotide 
in the transcribed RNA. The number of IRs in complete genomes
was first investigated with bioinformatics methods  
in long DNA sequences of eukaryotic (human and yeast) and bacterial 
({\it E.coli}) DNA \cite{Schroth}. Successive studies have
considered complete genomes such as the complete genomes of
eubacterium {\it Haemophilus influenzae} \cite{Smith}, archaebacterium
{\it Methanococcus jannaschii} and cyanobacterium
{\it Synechocystis} sp. PCC6803 \cite{Cox}. A Comparative genomic study 
of inverted repeats in prokaryotes has been investigated 
in Ref. \cite{Lillo}. 

An example of hairpin is shown in Fig. 1 for illustrative purposes. The figure
caption describes in detail all the different part of such secondary structure.
The example also presents a region of the stem with mismatches.

\begin{figure}[htbp] 
\centering{\resizebox{5cm}{!}{\includegraphics[scale=0.5]{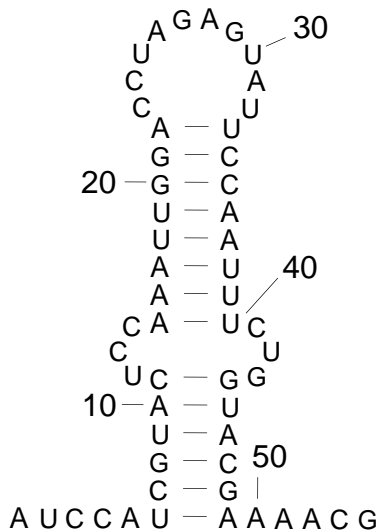}}}
\caption{An example of the hairpin structure formed by a single stranded RNA.
Bases from $6$ to $22$ constitute the left arm of the stem and bases
from $33$ to $49$ constitute the right arm. The loop is made by bases $23$ to
$32$. In this hairpin there is also a three base mismatch at bases $12$-$14$ and
$41$-$43$. In the terminology used in the paper the parameter of this inverted 
repeat are $\ell=17$, $m=10$, and $mis=3$.} 
\end{figure}

The simplest type of IR is the one without mismatches with the
additional condition that the base pair before and after 
the stem are not complementary base pairs.
Within this definition, Ref. \cite{Lillo} has shown that
the number of IRs expected in a 
genome under the simplest assumption of a random 
Bernoullian DNA is given by the equation      
\begin{equation}
n_{ex}(\ell,m)=N (1-2P_aP_t-2P_cP_g)^2 (2P_aP_t+2P_cP_g)^{\ell},
\end{equation}
where $N$ is the number of nucleotides in the genome sequence 
and $P_a$, $P_c$, $P_g$ and $P_t$ are the observed 
frequencies of nucleotides.
Eq. (1) shows that the number of expected IRs 
is independent of $m$ whereas it depends on the CG content
of the genome. The CG content vary considerably across different 
genomes and for long genomes also
across different regions of the same genome.

In the present study we are interested in a wider class of IRs in which the presence of 
mismatches is allowed, because these are present in many IRs with known biological role. 
Specifically, we detect all the IRs present in the complete genomes 
of viruses characterized by a stem length $\ell$ ranging from 6 to 20 and a loop length $m$
ranging from 3 to 10. Inside the stem up to 2 mismatches are allowed provided
that the number of links between complementary nucleotides inside the stem 
is always equal or larger than 6. With these constraints and with the additional
requirement of avoiding to count the same substructure as a portion of differently
classified structures we focus on three different classes
of IRs defined as follows. 
Example of the three different classes of 
inverted repeats detected are shown in the schematic drawing of Fig. 2.
The first one is characterized by a stem with no 
mismatches with the additional check that the three base pairs before and after 
the stem are not complementary base pairs (see scheme at Fig. 2(a)). 
The second one is a stem with 
one mismatch inside and  with the additional check that the two base 
pairs before and after the stem are not complementary base pairs (Fig. 2(b)). 
Finally the third
one is a stem with two mismatches inside and  with the additional condition 
that the base pair before and after the stem are not complementary base pairs
(Fig. 2(c)). In Ref. \cite{Lillo2} we generalize Eq. (1) for the three classes of structures considered here.


\begin{figure}[htbp] 
\centering{\resizebox{10cm}{!}{\includegraphics{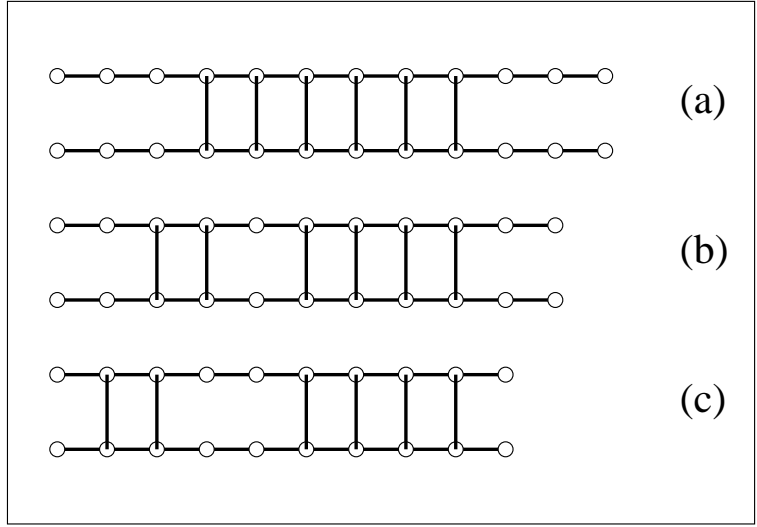}}}
\caption{Schematic classification of the stem of the 
three classes of inverted repeats 
investigated in the present study. Each horizontal line indicates one of the
RNA strands composing the stem of the potential secondary structure. 
A vertical line indicates the presence of a link between two complementary
nucleotides whereas the absence of a vertical line indicates no 
complementarity. Structures of the kind (a) have not mismatches. 
Structures of the kind (b) have 1 mismatch and of kind (c) have 2 mismatches.
They do not need to be contiguous.} \label{fig2}
\end{figure}

 These equations allow therefore to perform a statistical test of 
the null hypothesis that the number of detected IRs is compatible
with the assumption that viral genomes are Bernoullian 
symbolic sequences. This null hypothesis is equivalent to the 
assumption that IRs are observed in the genomes just by pure chance.

\section{$\chi^2$ tests}

We perform several $\chi^2$ test \cite{Test}. In all cases the $\chi^2$ test is performed 
by comparing the number of IRs of the three classes described in Fig. 2
for different structures. Each structure is defined by a single value of 
${\ell}$ ranging from 6 to 20. For each value of ${\ell}$, the loop length 
$m$ is varying from 3 to 10 and we verify the additional condition 
of observing at least 6 links within the stem. When the expected number 
of IR of a structure defined as before is larger than 5 we consider the 
structure of that kind as a degree of freedom of the $\chi^2$ test. 
When the number is smaller than 5 we aggregate different structures together 
until the number of expected IRs of the  aggregated structures is larger than 5. 
In our procedure, the test cannot 
be performed for a certain number of viruses not reaching the threshold value of 5 for at 
least one type of the three investigated structures. Due to the wide range of lengths of the investigated
genomes, the $\chi^2$ test is realized with a variable number of degree of freedom.
We set the confidence threshold of the $\chi^2$ test at 0.05. This implies that 
the null hypothesis is rejected when the p-value is smaller than 0.05. 
 
We have performed the $\chi^2$ test of the hypothesis that genomes are described 
by a Bernoullian sequence on 736 genomes of viruses. In this set, 324 genomes pass 
the test whereas in the remaining 412 (56 \%) the Bernoullian hypothesis must be 
rejected. In 409 cases out of 412 the statistical test is not 
passed due to an excess of the number of detected IRs. 

Having verified that the null hypothesis of IRs compatible with a Bernoullian 
DNA sequence is falsified in
a large number of viruses of our set, we have repeated the same test in the
coding and non coding regions of the genomes separately. Specifically, 
we label as coding regions all the regions of viral nucleic acids 
that are annotated in the databases as sequences coding for aminoacids in proteins.
We label as noncoding nucleic acid regions the remaining regions of the genomes, 
therefore including nucleic acid regions producing different kind of RNA.

When we investigate coding regions we are able to perform the test on 719 viruses. 
Within  this set 356 (50 \%) viruses do not pass the test whereas the remaining 
363 do pass it. In all 356 cases the statistical test is not 
passed due to an excess of the number of detected IRs. Moving to the noncoding regions, 
the test can be performed on 540 viruses. The lower number is due to the fact 
that noncoding regions are typically just 10 percent of the viral genomes and
therefore the number of expected IRs under the Bernoullian hypotesis is roughly 
a tenth of the number expected for the coding regions. Within  this set of 540 viruses, 
165 (31 \%) do not pass the test whereas the remaining 375 do pass it. In 162 cases 
out of 165 the statistical test is not passed due to an excess of the number of 
detected IRs with respect to the expected ones in term of the Bernoullian hypothesis.

We have therefore verified that the Bernoullian hypothesis saying IRs are
present in viral genomes just by chance is not passed in a significant fraction 
of the investigated viral genomes. Moreover, the tests performed separately in coding and noncoding regions
show that the excess of IRs is observed both in coding and noncoding regions.

\begin{figure}[htbp]
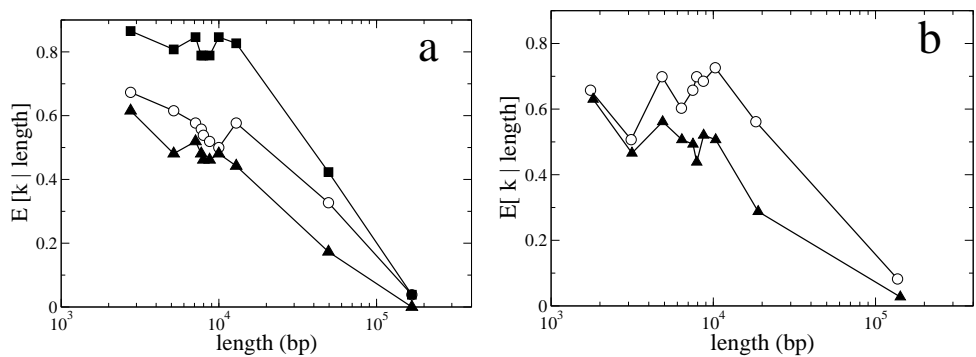
 
              {\hbox{
                {\includegraphics[scale=0.26]{Fig3a.eps}}
              \hspace{0.1 cm}
                {\includegraphics[scale=0.26]{Fig3b.eps}}
              }}
              \caption{In panel a) it is shown the mean value of the 
              parameter $k$ describing if the 
viral genome passes the Bernoullian test ($k=1$ when $p>0.05$) or 
the opposite case ($k=0$ when $p<0.05$) conditioned to the 
length of the viral genome. The investigation is performed for
a set of 524 viruses where the statistical test can be performed 
in the entire genome (triangles),
its coding (circles) and noncoding (squares) regions. 
Analogously, panel b) shows the results of the same kind of analysis 
performed in the entire genome of 736 viruses under the assumption of a Bernoullian null hypothesis (triangles) and 
of 738 viruses under the assumption of a Markovian null hypothesis (circles). Each symbol 
groups the same number of viruses.} \label{fig3}
\end{figure}

To shed light on the parameters influencing the validation or falsification 
of the null hypothesis in our viral database, we have investigated the 
number of viruses passing the Bernoullian test 
conditioned to the length of the viral genome. In this investigation we aim to
check the characteristic of viruses in the entire genome and both in the 
coding and noncoding regions. For this reason we have selected the viruses
where tests are possible both in the coding and in the noncoding regions. 
This set is composed of 524 viruses. Results are summarized in panel a) of Fig. \ref{fig3}.
In the figure we show the value of the mean value $E[k|length]$ 
of the parameter $k$ conditioned to the length of the viral genome. The parameter 
$k$ assumes the value $k=1$ if the genome passes the Bernoullian test 
or the value $k=0$ in the opposite case.
The three curves obtained respectively for the entire genome (triangles),
its coding (circles) and noncoding (squares) regions share
the same global behavior. The percentage of viruses passing the test 
is higher for shorter viruses and this percentage is steeply declining 
for length longer than 30,000 nucleotides. The conditional mean value
reaches approximately the zero value for the longest genomes. 
The figure also shows that at fixed value of the genome length the 
test is passed with a higher percentage in noncoding than in coding regions.
Global exceedence of the number of 
detected IRs with respect
to the number of IRs expected in terms of a Bernoullian hypothesis is 
detected in a large number of viruses both in coding and in noncoding 
regions. The exceedence is progressively more pronounced in longer viral genomes. 

The results we have obtained are not due to the fact that Bernoullian
hypothesis is a zero--order null hypothesis not well reproducing the
statistical properties of DNA and RNA sequences. To prove this sentence
we have repeated the test by comparing the number of detected IRs 
in each complete genome with the
number of IRs observed in a numerically simulated genome generated according 
to a 1-order Markov chain. The expected number of IRs is 
computed by simulating 100 different realizations of each genome
with the same measured Markovian transition matrix.
We have first measured the empirical Markovian transition matrix 
and therefore performed the test in all viral genomes.
The $\chi^2$ test with the Markovian hypothesys is performed on 738 viral genomes. Within 
this set of 738 viruses,  309 (42 \%) 
do not pass the test whereas the remaining 429 do pass it. In 306 cases 
out of 309 the statistical test is not 
passed due to an excess of the number of detected IRs.
This result shows that the Bernoullian assumption does not
give results too different from the more accurate Markovian one. Panel b) of 
Fig. \ref{fig3} shows the mean of the $k$ parameter conditioned 
to the length of viral genomes for the Bernoullian and the Markovian hypotheses.
The investigated sets comprises 736 viruses for the Bernoullian case and 738 
for the Markovian one. In the rest of this paper we will present results obtained by 
using as a testing assumption the Bernoullian model. This 
choice is motivated by the fact that under the Bernoullian assumption 
we are able to use an analytical estimation of the expected number
of IRs for each kind of the investigated structure whereas the 
expected number of IRs under the Markovian assumption can be obtained
only on a statistical basis by performing numerical simulations.

The next step is then to investigate how uniform is the localization of 
each kind of structure in each virus with respect to the kind
of considered coding or noncoding region. This new test is done 
separately for each structure identified as type a, b or c structure
according to the classification of Fig. \ref{fig2} and by its stem length $\ell$.
The null hypothesis 
is done in terms of a Bernoullian sequence to take advantage 
of the knowledge of analytical relations for the number of IRs expected
for each structure. The test is devised to evaluate the degree of uniformity
of the localization of observed structure of IRs inside each genome.
The estimation of the number of expected IRs in coding $N^{cr}_{uni}(t,\ell)$ and 
noncoding $N^{ncr}_{uni}(t,\ell)$ regions under the assumption of not preferential (uniform)
localization is done by using both the information about the total number 
of detected structures in the genome $N_{obs} (t,\ell)$ and
the frequencies expected in terms of the Bernoullian hypothesis through the equations
\begin{equation}
  N^{cr}_{uni}(t,\ell)= 
  N_{obs} \frac{N^{cr}_{Ber} (t,\ell)}{N^{cr}_{Ber} (t,\ell)+N^{ncr}_{Ber}(t,\ell)}, \label{this}
\end{equation}
and
\begin{equation}
  N^{ncr}_{uni}(t,\ell)= 
  N_{obs} \frac{N^{ncr}_{Ber} (t,\ell)}{N^{cr}_{Ber} (t,\ell)+N^{ncr}_{Ber}(t,\ell)}, \label{that}
\end{equation}
where $t$ indicates the type of the structure (a, b or c) and $\ell \ge 6$ the stem length. 
These equations take into account the possibility that coding and noncoding regions might have different nucleotide frequencies.

We are able to perform the test in 1316 structures of 524 different viruses. 
Among these structures 1070 of 423 distinct viruses are consistent with the assumptions 
used in the test, which are (i) no preferential 
location between coding and noncoding and (ii) frequency of the IRs proportional 
to a Bernoullian expectation. Only in the remaining 246 (19 \%) structures of 104 distinct viruses
the statistical test is not passed and therefore in this restrict number of cases
there might be a preferential location
of these structure in one of the two consider regions. By looking at the specific
contributions to the $\chi^2$ test we note that in 204 cases of the 246 considered the
detected structures are preferentially located in the noncoding regions.
At first sight this result can be seen as not consistent with the 
results summarized in Fig. 3a where the Bernoullian test is more easily passed in noncoding
rather than coding regions. But indeed there is no contradiction. In fact, among the 104 viruses having the 246
structures that do not pass the Bernoullian test and show a preferential
location in the noncoding regions, 79 of them are longer 
than 10,000 bp. Moreover, these 79 viruses contains 220 of the 246 
considered structures. For viruses of such length 
the difference between the conditional mean value of the test indicator
$k$ for coding and noncoding regions is much less pronounced than 
for viruses shorter than 10,000 bp and tend to disappear for longer 
viruses. The last $\chi^2$ test is preferentially passed by structures 
located in longest viruses.

\section{Conclusions}

The present study has sistematically investigated the presence of IRs
in 738 complete genomes of viruses. 
The investigated IRs can be both with
a perfect matching of links within the stem and with the presence
of mismatches up to a maximal value of 2. The empirically observed 
inverted repeats are compared with the values expected in terms of
Bernoullian or Markovian model of genomes.

We find as a statistical regularity that the number of observed 
IRs are often greater than the one expected in terms of a
Bernoullian or Markovian model in several of the viruses
and in almost all those with a genome longer than 30,000 bp.

There is not a pronounced preferential location of these IRs in coding
or noncoding regions in the majority of the considered viruses. 
This result is different from the one obtained by investigating
complete genomes of bacteria \cite{Lillo} where a distinct preferential
location of long structure of IRs in noncoding regions was observed. 

We have therefore devised a methodology to detect sets of IRs whose existence 
cannot be explained in terms of simple random models of genome
sequences. The selection of these sets of IRs may allow the future analysis
of these secondary structure finalized to several distinct goals such as, 
for example, the detection of the degree of homology
among them both in terms of sequence similarity or in terms of free energy
of the secondary structures.

\section*{Acknowledgements}
This work has been funded by the MIUR project 
``A new approach to drug design. From quantum mechanics 
to the screening of anti-viral drugs". M.S. acknowledges a fellowship within this project. 
S.M. acknowledges support
from FIRB project ``Cellular Self-Organizing nets and 
chaotic nonlinear dynamics to model and control complex system".

\end{document}